\documentclass[letterpaper,twocolumn,10pt]{article}
\usepackage{usenix2022_SOUPS}

\usepackage{tikz}
\usepackage{amsmath}

\usepackage{graphicx}
\usepackage{multirow}
\usepackage{caption}

\usepackage{comment}
\usepackage{array}
\usepackage{lipsum}
\usepackage{caption}
\usepackage{subcaption}
\begin{document}
%-------------------------------------------------------------------------------

%don't want date printed
\date{}

\title{\Large \bf What You See is Not What You Get: \\The Role of Email Presentation in Phishing Susceptibility}

% if you leave this blank it will default to a possibly ugly attempt 
% to make the contents of the \author command below into a string
\def\plainauthor{Author name(s) for PDF metadata. Don't forget to anonymize for submission!}

%for single author (just remove % characters)
% \author{
% {\rm Anonymized Authors}\\
% }

\author{
{\rm Sijie Zhuo}\\
University of Auckland
\and
{\rm Robert Biddle}\\
University of Auckland
\and
{\rm Lucas Betts}\\
University of Auckland
\and
{\rm Nalin Asanka Gamagedara Arachchilage}\\
University of Auckland
\and
{\rm Yun Sing Koh}\\
University of Auckland
\and
{\rm Danielle Lottridge}\\
University of Auckland
\and
{\rm Giovanni Russello}\\
University of Auckland
% copy the following lines to add more authors
% \and
% {\rm Name}\\
%Name Institution
} % end author

\maketitle
\thecopyright

%-------------------------------------------------------------------------------
\begin{abstract}
%-------------------------------------------------------------------------------
Phishing is one of the most prevalent social engineering attacks that targets both organizations and individuals. It is crucial to understand how email presentation impacts users' reactions to phishing attacks. We speculated that the device and email presentation may play a role, and, in particular, that how links are shown might influence susceptibility. Collaborating with the IT Services unit of a large organization doing a phishing training exercise, we conducted a study to explore the effects of the device and the presentation of links. Our findings indicate that mobile device and computer users were equally likely to click on unmasked links, however mobile device users were more likely to click on masked links compared to computer users. These findings suggest that link presentation plays a significant role in users' susceptibility to phishing attacks.
\end{abstract}

%-------------------------------------------------------------------------------
\section{Introduction}
%-------------------------------------------------------------------------------
Phishing attacks have become a common method of cyberattack in recent years. Even with the existence of phishing filters and countermeasures, the end users are still the last line of defense against such attacks. Once a phishing email passes all the technical defenses and reaches the users' inbox, the users themselves have to make the right judgment. Users' phishing susceptibility can be influenced by many factors, including their level of education, knowledge and experience. Their ability to detect phishing emails and making good decisions is also influenced by their awareness, mood, beliefs and involvement in reading the email. The characteristics of the email can also influence users' decision-making \cite{zhuo2022sok, norris2019psychology, franz2021sok}.

Various human-centered solutions have been developed to help reduce users' phishing susceptibility. Phishing training is one of the most common approaches to educating about phishing and how to prevent it. Different training tools have been developed to help users identify phishing cues, including text-based materials, video-based training \cite{volkamer2018developing}, and game-based training solutions \cite{sheng2007anti, wen2019hack}. There is no doubt that phishing training can reduce users' susceptibility to phishing \cite{baillon2019informing, lain2021phishing}, but periodic follow-up training is then necessary to help them maintain the level of knowledge and awareness \cite{jampen2020don, reinheimer2020investigation}. In practice, when checking emails, users usually focus on the email content and treat the emails as legitimate by default. Only when some cues in the email make them feel suspicious, do users change their mindset to validate the legitimacy of the email \cite{wash2020experts}. When reading emails, what the user sees is how their email client renders the email. We speculate that email clients also play a role in influencing users' phishing susceptibility.

The device used for checking emails may impact the email's presentation and how users interact with it. One common difference between computer and smartphone involves links. Phishers usually manipulate the phishing link to look trustworthy; they often hide the link behind a button or text to ensure their phishy URL is less visible to the user. On computers, users can easily view the landing page URL of an embedded link by hovering over it. However, on smartphones, users have to tap and hold the link to view the URL. This interaction is less familiar and more complicated for the users, making them more likely to click on phishing links directly, and thus take the first step in being phished. Further, the limited physical size of smartphones leads to certain design choices in email clients, such as hiding certain details of the email (e.g. the sender's email address and utility function buttons), which may bias users' judgment. These issues suggest that users may be more susceptible to phishing when using mobile devices.

In this paper, we report on a study conducted with the IT service unit of a large organization while it conducted a regular phishing training exercise. The study had two main aims: a) to investigate whether the use of different devices can influence users' tendency of clicking on phishing links; b) to study whether the visual presentation of phishing links influences users clicking behavior and thus their susceptibility to phishing. To the best of our knowledge, we are the first paper that explores these topics on a large scale. 

The rest of this paper is organized as follows. In Section \ref{RelatedWork}, we provide relevant background on the influence of UI design on users' behavior, and the differences in UI design of existing email clients. In Section \ref{Methodology}, we explain our methodology, and how our study was conducted as part of a large-scale phishing training exercise. In Section \ref{Result}, we present the results of our study and our analyses. Then in Section \ref{Discussion}, we discuss our findings and suggest the implications. Lastly, in Section \ref{Conclusion}, we present our conclusions and suggest future directions.

\section{Related Work} \label{RelatedWork}
%-------------------------------------------------------------------------------

This section first recalls evidence from the research literature on how user interface (UI) design can influence user behaviors, and how design in email clients can affect security risks. Then we discuss the differences between computers and mobile devices for checking email, including the differences in the devices' physical characteristics, interaction methods, and users' security awareness. Finally, we give an overview of past research on the impact of phishing link presentation on user behavior and susceptibility to phishing.

\subsubsection*{Influence of UI design} Many studies have shown that UI design can influence users' behavior and choices \cite{schneider2018digital, rendell2021nature,lau2018influence, alves2020incorporating, anwar2020impact}. For instance, Schneider et al.'s research \cite{schneider2018digital} shows that UI designers can nudge users into making certain choices unconsciously through the designs of interfaces. These nudging techniques have been used widely in commercial websites to attract customers. For example, one can increase the attractiveness of an option by placing it next to an unattractive option. 

UI designs not only influence users' selection of options, but also their perceptions and attitudes toward the interfaces. Both Anwar et al.'s study \cite{anwar2020impact} and Rendell et al.'s study \cite{rendell2021nature} show that an appropriate selection of images, styles, and color schemes will encourage users' engagement with the website. In the example of online shopping websites, this would greatly increase users' purchase intentions and revisit intentions. In addition, Stojmenovi\'{c} et al. show that visual appeal influences the perception of security \cite{stojmenovic2022beauty}. Users are more likely to trust a website that is beautifully designed.

The impact of UI design on users can differ from individual to individual. Alves et al.'s review \cite{alves2020incorporating} shows that users with different personalities have different preferences for UI designs and aesthetics, and these differences can affect their task performance and information-seeking. It has been found that users tend to be more efficient when the interface is designed to match their personality \cite{kostov2001development}.

The impact of UI design can also apply to email clients. Users' security awareness and ability to detect phishing emails can be improved by manipulating the UI elements in the email client. Anderson's study \cite{anderson2016users} explored the design of phishing warnings and shows that users gradually ignore warning messages after several occurrences. The use of polymorphic warning messages can slow down the forming of such habituation, which can help users stay alert longer. Petelka et al.\cite{petelka2019put} modified the embedded links in emails to help users pay more attention to the landing page URL. In their study, the most effective approach was to deactivate the link, and force the users to interact with the raw URL in the hover box to reach the landing page. These studies demonstrate the potential in improving users' security performance through changes in email client design.

\subsubsection*{Computers vs. mobile devices} In recent years, the ubiquity of smartphones and tablets has led to a rising trend of using them for checking emails \cite{mobileStats}. Due to the differences in functionality and characteristics, the interaction mode is different when using different devices. For instance, users interact with smartphones by directly tapping and swiping on the screen. In contrast, when using a computer, the interaction is done by controlling the cursor or typing keys on the keyboard. Besides, smartphones have smaller screens than computers, meaning smartphones can only display a limited amount of content at one time. When designing email clients for mobile devices, designers need to carefully choose important content to be displayed on the small screen, meanwhile supporting utility components for interaction and usability. This sometimes leads to hiding some details, such as the sender's full email address. From the security perspective, hiding the email address means that users would have one fewer cue which might help them identify malicious emails, and thus would lead to higher phishing susceptibility.

Moreover, the mobility of smartphones means that they are often used during casual situations or outdoor environments, which usually involve more distractions than using a computer. User performance can worsen in a distracting environment, since users have to spend additional working memory to suppress the distractions so that they can focus on their tasks \cite{barrett2004individual}. When checking emails, less cognitive effort and attention spent on the email may lead to careless behaviors and a lower chance of detecting phishing emails, thus under higher risk of falling for phishing \cite{norris2019psychology, zhuo2022sok}.

Research has found that users' mindsets when using different devices are different. Breitinger et al.'s study \cite{breitinger2020survey} shows that users are more concerned about protecting their smartphones from attackers getting physical access, so they tend to focus more on password security and messaging security when using their phones. Whereas for their computers, they tend to focus more on secure detection and virus scanners to protect their data and privacy. Further, users tend to have lower information security awareness (ISA) when using smartphones than computers \cite{breitinger2020survey, mcgill2017old, bitton2018taxonomy, koyuncu2019security, felt2011phishing}. Many users install security software on their computer, but only about one-third of users may consider it as necessary for their smartphone \cite{breitinger2020survey}.

Most phishing-related training primarily focuses on computer users, including training materials and solutions. Thus, educated users usually have the mindset of how to deal with emails when using their computer, such as hovering over links before clicking. However, there is a large population of users who check their email using their smartphones or tablets. Due to the lack of specific training targeting mobile device users, these users could have less willingness and knowledge of how to view the URL of the link without opening the link. 

\subsubsection*{Visual presentation of phishing links} There has been some research on phishing email visual presentation. For example, users are more likely to trust emails with professional-looking visual presentation, including the presence of logos, copyright statements and other seemingly authentic cues \cite{williams2019persuasive, blythe2011f}. The same phishing message may display differently across clients because email clients have different ways of presenting messages. Some email clients have settings that disallow downloading of images, and in this case, logos are not shown to users. The presentation of an email in a desktop email client may be different from a mobile device email client because of the screen size. 

%Some email clients may disallow picture download by default, so users are not able to view the full content unless they manually changed the settings or allow picture download. In this case, the presence or absence of logo does not necessarily help user decide the legitimacy of the email. 

The phishing link is one of the most important parts of phishing mail. For attackers, users clicking on the phishing link can be considered as an indicator of the potential success of phishing, and the link may allow the attacker to identify the email address of the user. The attractiveness of the phishing link plays an essential part in the success of the attack. There have been studies that focus on identifying techniques that have been used to manipulate phishing URLs to make them look legitimate to users \cite{garera2007framework, althobaiti2021don}, and studies on developing tools to help users pay attention to links \cite{volkamer2017user}. Attackers will often not display the full phishing URL directly in the email. The link is usually masked with text or buttons, so unless users hover over the link on a desktop or tap and hold the link on a smartphone, they will not see the actual phishing URL. 
% The knowledge of detecting phishing URLs will be useless if users base their judgment on the presentation of the email and phishing link. Therefore, the visual presentation of the phishing link can help us better understand why users fall for phishing.

%-------------------------------------------------------------------------------
\section{Method} \label{Methodology}
%-------------------------------------------------------------------------------
In this section, we first introduce the research questions and our hypotheses. Then we explain our study setting, the materials used, and the data collection process.

\subsection{Research Questions and Hypotheses}

Our study aims to answer two research questions. The first one is \textbf{RQ1: is user phishing susceptibility influenced by the devices they use when checking emails?} Our hypothesis is \textit{
H1: Users are more likely to click on phishing links when using email clients on mobile devices than on computers.} 

The second research question is \textbf{RQ2: Does the presentation of email phishing links influence users' phishing susceptibility?} We hypothesize that \textit{
H2: Users are more likely to click on phishing links when masked with buttons or hypertext than when shown as URLs.} 

Both buttons and hypertext are common in websites and emails. Both can be designed to look more visually attractive to the user than raw URLs. But in general, there is more flexibility in designing the visual presentation of a button than hypertext, hence users may be more likely to distinguish a button from the surrounding non-link content than hypertext. Also, buttons have more affordance for pushing and clicking, which may further motivate users to click. Hence, we formed an additional last hypothesis:
\textit{H3: Users are more likely to click on phishing links if they are hidden with HTML buttons, compared with being hidden with hypertext.}

Our data is categorical, indicating counts of users opening the email, clicking on links, and on which devices. We therefore use chi-squared tests to determine the significance of differences, and Pearson's phi ($\phi$) coefficient for effect size; our tests have one degree of freedom, so a value of .1 is considered a small effect size, .3 is moderate, and .5 is large\cite{hemphill2003interpreting}. Our hypotheses are directional, and with the 2x2 structure of the contingencies we can use the chi-squared residuals to test whether any difference is in the direction we hypothesize.

\subsection{Study Setting}

In order to understand how users realistically interact with phishing emails, we need to study how users react to real emails received via their real email addresses. To do this, we collaborated with a large organization that regularly conducts cybersecurity training and phishing simulations to educate employees. The organization's IT services unit sends a simulated phishing email to all their employees, where those clicking on a simulated phishing link are directed to a webpage with training material about phishing.  

We collaborated with the IT services unit to conduct our study within their training exercise. We collaborated with the organization's cybersecurity team and scheduled the training for November 2022, in which they sent out one phishing email to each staff member in the organization. We helped the team design the phishing email templates and the landing pages. At the end of the campaign, we received anonymous data from the cybersecurity team for further analysis. Our study protocol was reviewed and approved by our Human Participants Ethics Committee and by the organization's Chief Information Security Officer.

Before the actual campaign, we pilot tested the phishing simulation with 12 members of the organization's cybersecurity team. We then tested within the whole IT service unit of 374 users to test with a larger, more diverse group. All this was to pilot test our process, and we do not present any data for these steps in the results. 
The main campaign was sent to over 12,000 employee email addresses in the organization.

%\subsection{Participants}

% TODO, discuss that the low number of open could be due to sending emails to inactive accounts, formar employee, some users may not know they been given a account, and 
%In this organization, employees would automatically receive one email account (Outlook Exchange account) when join the organization. However, many employees never used the given email account or did not know that they have been given an account. Due to the nature of the organization, we are not able to distinguish between active and inactive users, so we decided to send out phishing emails to all accounts.

% The two campaigns are targeting the same users, the organization staff's working email accounts, which are all Outlook Exchange accounts. 

\subsection{Study Materials}

To study the second research question, we crafted three variations of a phishing email using the same sender address, subject, and main text. The email claims that the user has waiting email messages that must be checked and released from an online system. This mimics notifications sometimes sent in this organization. The organization is using the NIST phish scale \cite{steves2020categorizing} in phishing training exercises to assess the detection difficulty of phishing emails, which limits our choices of phishing emails and difficulty. For this campaign, we were asked by the IT service unit of the organization to use a phishing email template with a detection difficulty of moderate difficulty.

The three versions differ only in the visual presentation of the phishing link, as follows:
\begin{itemize}
    \item Version 1 (raw URL): Display the real (simulated) phishing link ``https://foxypdf.nzs.net" as clickable in the email. Note that we have slightly modified the domain name of the URL to keep the organization anonymized.
    \item Version 2 (button with text): Display the phishing link as a button, with the text ``Release held emails". In HTML, this is done by placing a button tag inside a hyperlink tag.
    \item Version 3 (hypertext): Display as the clickable anchor text ``Release held emails", instead of the real URL.
\end{itemize}

\begin{figure}[h]
     \centering
     \begin{subfigure}[b]{0.47\textwidth}
         \centering
         \includegraphics[width=\textwidth]{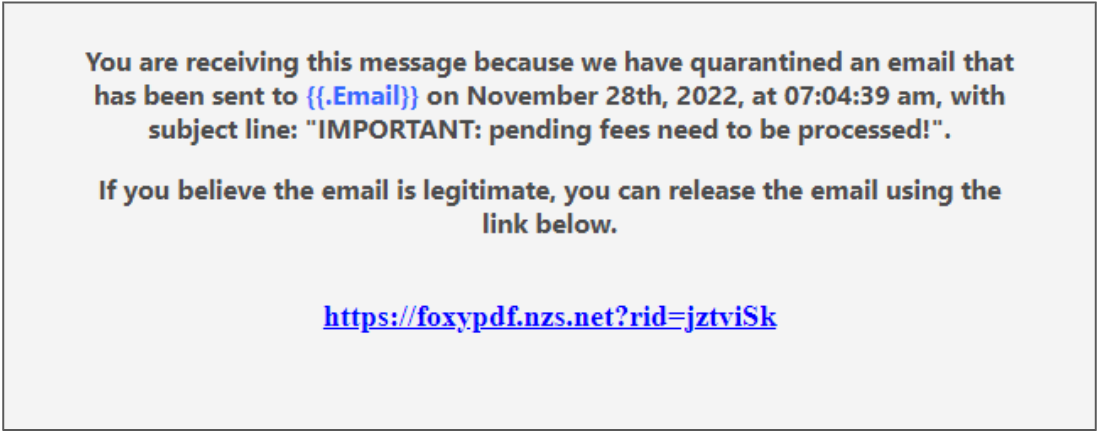}
         \caption{Raw URL version}
         \label{fig:p_raw}
     \end{subfigure}
     \hfill
     \begin{subfigure}[b]{0.47\textwidth}
         \centering
         \includegraphics[width=\textwidth]{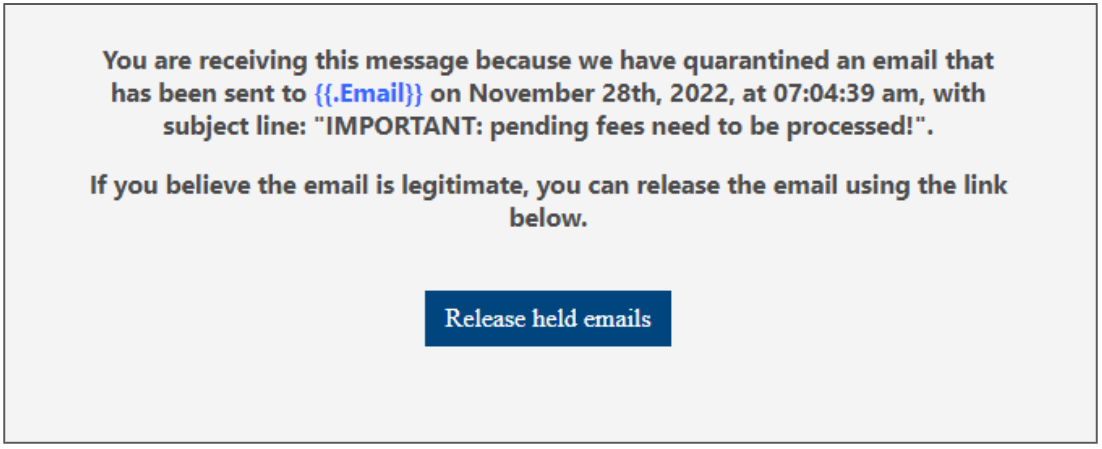}
         \caption{Button version}
         \label{fig:p_button}
     \end{subfigure}
     \hfill
     \begin{subfigure}[b]{0.47\textwidth}
         \centering
         \includegraphics[width=\textwidth]{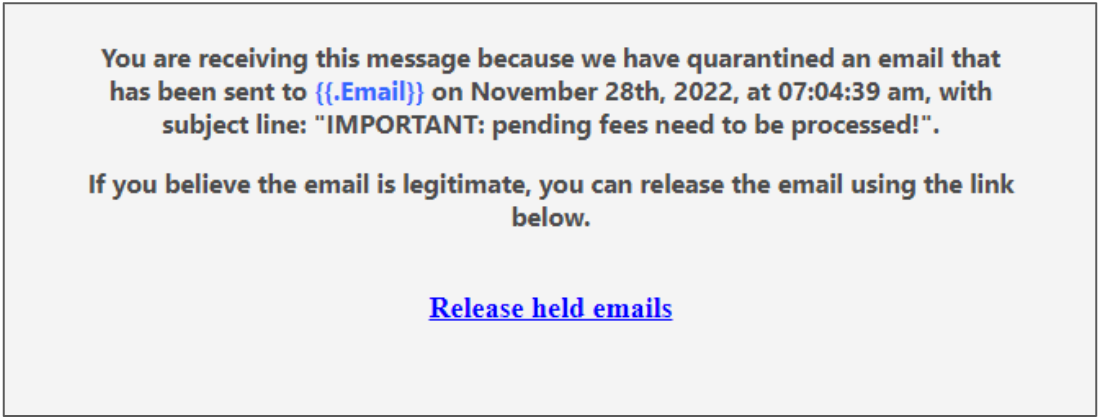}
         \caption{Hypertext version}
         \label{fig:p_hypertext}
     \end{subfigure}
        \caption{The three campaign emails: raw URL, button, and anchor text}
        \label{fig:email3}
\end{figure}

In Figure \ref{fig:email3}, we show how the three versions of the email appeared. In all 
three versions, the header details were the same. The sender was: \textbf{Security Gateway Notification <security-gateway@onlinesecurity.net>}, and the subject line: was \textbf{You have new on hold emails}. The actual URL and landing page were the same. The landing page itself, if users did open it, explained that this was a phishing simulation, and led to phishing education material.

In the campaign, we separated the users randomly into three groups, and members of each group received one of the variations of the phishing email. 

\subsection{Data Collection}

Working with the organization's cybersecurity team, we managed the simulated phishing email and collected data, using the open-source phishing framework GoPhish\footnote{GoPhish, Open-Source Phishing Framework by Jordan Wright: \url{https://getgophish.com/}}. The GoPhish installation was installed and managed by the cybersecurity team.

Each email address used with GoPhish is given a unique anonymous ID. GoPhish then records three types of events: email sent, email opened, and link clicked. The email opened event data is collected by attaching a tracking image in the email that requires access to the GoPhish server, and the tracking image can be related to the tracking ID. The link clicked event is generated by the user clicking on the link in the email (via the raw URL, or button, or anchor text) that contacts the GoPhish server, and again can be related to the tracking ID. For each of these open and click events, GoPhish is able to record the timestamp of the event, the IP address, and the user-agent string. The user-agent string can then be used to identify the application, operating system, vendor, and/or version of the application that sends the user-agent request \cite{useragent}. 

The data produced by GoPhish was first pre-processed to remove non-user actions. Due to the complexities in the email transferring system and the use of different email systems, some open/click attempts are generated before reaching the recipient's inbox, which creates additional noise in the data. Some email clients (such as Gmail and Apple Mail) route their emails through proxy servers or caching servers before the users receive the emails, which can generate open/click events that are not performed by the user. We filtered this noise by checking the IP address associated with the events, and removed the ones that came from hosting service companies. If the Autonomous System Number (ASN) of the IP address belongs to the organization of the email client or some hosting service companies instead of the user's local network provider, we determined it is likely that this event was not actually the result of the user activity. 

Both the open and clicked events recorded by GoPhish include user-agent data, but the click event will indicate the client software, typically a web browser, used to display the landing page. For our analysis, we therefore used the user-agent string of the open event when classifying the platform used both to open the email, and to click the link: typically an email client.

The image tracking system used by GoPhish to determine email openings is not perfect. As mentioned earlier, some email clients allow users to block images within emails, either optionally or by default. As also mentioned above, email systems (laudably) include measures to obfuscate email access in other ways. Moreover, user-agent strings are not always reliable, and are sometimes made vague for privacy and security reasons. We acknowledge that these issues constitute limitations in our study. However, we have no reason to believe that users who block images or are impacted by other security measures would be differently distributed across our three email version groups. We therefore propose that while our overall numbers may be affected, the differences in our conditions are worthy of notice.

%The user-agent string of the click attempts generated by GoPhish are associated with the software that opened the landing page, instead of the email client that brings user to the landing page. Therefore, to understand how the user read the email before click on the link, we replaced the user-agent strings with the user-agent string of the user's last open attempt.

% pixel problem, will not effect our result

%-------------------------------------------------------------------------------
\section{Results} \label{Result}
%-------------------------------------------------------------------------------
\subsection{Campaign Overview}

The campaign started on a workday Monday at noon, and emails with sent to 12,639 email addresses. We received data until Wednesday afternoon, for a duration of just over two days. The data showed that between two and three thousand users opened the email. This number is much lower than the number of emails sent for several reasons. While the organization has several thousand current full-time employees, it provides email addresses to many part-time staff, and even maintains email addresses for former staff, and many of these people likely never read their emails. It is also possible some users may have seen the subject line and simply ignored the email as irrelevant. Moreover, we only received data for about two days, so there will have been users who opened the email after we received the data. Since most users who open email do so within the first few hours after receiving \cite{oest2020sunrise}, we would only expect a small number of activities after two days, so we believe our results were not overly affected. This bounded timeline is utilized by the IT security team in order to scope the workload of the IT support teams who receive reported phishing emails and questions from employees. 

% TODO explore on the effect size, even though the effect size is small, but to get a large effect size, the differnce in click rate has to be hugely different. so we will say that even though the effec size in the study is not big, it still reflects the effectiveness of the manipulation

% may want to find more convincing explanation regarding the browsers
We first inspected the data for unusual activity, as we expect usual user behavior to be at most a few clicks from a single email client. Thus we removed 27 users from the dataset who performed more than 20 interactions (including both email open activities and link click activities), and 22 users who clicked on the phishing link from multiple email clients; both these made it difficult to determine a single device and sequence of behavior.  %We interpret these unusual behaviors to have been generated by GoPhish or performed by the IT security team members.  %Most users would not read their email multiple times, and at least would not perform more than 20 interactions (open and click). We feel these behaviors are rare and may skew our data. For example, for the users who clicked using multiple devices, we are uncertain which cues influenced their behavior because the email client UI may be different for different devices. 
We also became uncertain of presentation issues relating to ``web mail'' clients, i.e., email clients running in a web browser. In particular, we learned that the organization configured their web mail service's default setting to disallow the automatic downloading of images, which prevented valid counts of email opening. We therefore decided to focus on email-specific application software, and filtered out results where email was opened within web browsers.
% The organization  configured their \textcolor{red} {webmail} service's default setting to disallow the automatic downloading of images, which prevented  valid counts of user interactions with webmail clients. Thus  we removed the users who opened their emails in web clients and our results focus on email client applications.
%There were users who manually changes the automatic picture download policy --- this group of users are small ---- we observed less than 20 users using browsers with images. 
%\textbf{DANIELLE: DOES THIS SOUND OK?} 

%\textbf{ROBT: MAYBE WE SHOULD JUST SAW WE DID NOT WANT TO STUDY BROWSER EMAIL: THE EXPLANATION SOUNDS WEAK.} 
%We also removed the users who opened their emails in web clients because the organization has configured the email service's default setting to disallow the automatic downloading of images, which influenced our data collection. From our observation, the change was not applied to email client applications; thus we focus our study on email client applications. We acknowledge that there will be users who manually change the automatic picture download policy, but since this group of users is small enough (observed less than 20 users using browsers), that will not skew the result. 

After filtering, the data showed 2,375 users who opened our email, about 20\% of the total number sent. Of these, 285 clicked on the link, so about 12\% of those who opened the email. These users generated 3,921 open actions, so many opened the email more than once, and there were 304 click actions, which means a few people clicked more than once using the same platform. 

\subsection{Device Use}

By processing the user-agent string of each event, we extracted the platform, operating system and email client used to perform the event. We grouped our data based on the device used for viewing the email: computers, mobile devices, and unclassified. The unclassified category mainly contains users who use Apple Mail and other email clients where it is not possible to identify the platform based on the user-agent string. Due to Apple Mail's mail privacy protection mechanism, the device used to open the email is not identified \footnote{https://www.apple.com/legal/privacy/data/en/mail-privacy-protection/}. From our observation, all platforms (Mac, iPad or iPhone) use the same user-agent at the time of writing ("Mozilla/5.0"). The distribution of the groups is shown in Table \ref{tab:campaign1_devices}. The overall total count is slightly larger than the number of users who opened the email, this is because some users can open the email using multiple devices and email clients, but since these users at most clicked on one type of email client, our data is not affected.

\begin{table}[]
\centering
\caption{Distribution of the campaign users' clicking actions using different devices}
\label{tab:campaign1_devices}
\begin{tabular}{|p{1.4cm}|>{\raggedleft\arraybackslash}p{1.4cm}|>{\raggedleft\arraybackslash}p{1.0cm}|>{\raggedleft\arraybackslash}p{1.6cm}|r|}
\hline
                           & \textbf{Computer}  & \textbf{Mobile}  & \textbf{Unclassified}   & \textbf{Sum} \\ \hline
\textbf{Users clicked}     & 88                 & 135                           & 62                & 285           \\ \hline
\textbf{Users not clicked} & 538                & 591                           & 1200              & 2339         \\ \hline
\textbf{Sum}               & 626                & 726                           & 1262              & 2614         \\ \hline
\end{tabular}
\end{table}

\subsubsection{H1 Testing}

The data allows us to test Hypothesis H1: \textit{Users are more likely to click on phishing links when using mobile devices than on computers.}

A chi-squared test was performed to examine if there was a connection between users link clicking behaviors and the device used for checking emails. For this chi-squared test, we removed the unclassified group because of the uncertainty of which kind of device was used. The result was significant, $\chi^2 (1,  N = 1352) = 5.02, p = .025$ (Pearson residual: $+1.4$). Here, and below, we give the residual for clicks in the first stated condition -- mobile devices, in this case. The positive value shows that mobile devices had more than the expected number of clicks were the categories the same. We thus conclude H1 is supported, and user behavior is significantly different when using different devices for checking emails. However, we calculated the phi coefficient for effect size as $\phi = .06$, showing only a very small effect, so we conclude that the difference in clicking behavior attributable to the device used is quite minor.

These results are also shown in the first row of Table \ref{tab:campaign1_devs}.

%\subsubsection{Device and Link Presentation Testing}
We further carried out tests to explore whether the use of different devices might have an influence on users' clicking behavior. We compared the data for mobile devices and for computers, looking at each link presentation group. For the raw URL, the result was: $\chi^2 (1,  N = 396) < .01, p = .93$ (Pearson residual: $-0.1$), so no significant difference. In other words, when the raw URL is presented, the device does not matter.

For the button group, the result was: $\chi^2 (1,  N = 371) = 5.39, p = .02, \phi = .12$ (Pearson residual: $+1.3$), and similarly for the hypertext group, it was: $\chi^2 (1,  N = 585) = 5.98, p = .01, \phi = .10$ (Pearson residual: $+1.6)$). These results broadly align with our findings above, that it is link presentation, more than device type, that shows different behavior. These results are also shown in Table \ref{tab:campaign1_devs}.

\begin{table}[h]
\centering
\caption{Click behavior for mobile devices vs. computers:  chi-squared test results; residual shown for mobile device clicks}
\label{tab:campaign1_devs}
\begin{tabular}{|p{2.2cm}|>{\raggedleft\arraybackslash}p{1.2cm}|>{\raggedleft\arraybackslash}p{1cm}|>{\raggedleft\arraybackslash}p{1.3cm}|}
\hline
\textbf{Condition}    & \textbf{$\chi^2(1)$ p-value}  & \textbf{Effect size $\phi$} & Pearson Residual\\ \hline
All click behavior& .025 & 0.06 & +1.4 \\ \hline
Raw URL click behavior               & .93           & $<$ .01   & -0.1    \\ \hline
Button click behavior              & .02           & .12   &  +1.3    \\ \hline
Hypertext click behavior           & .01           & .10     &  +1.6   \\ \hline
\end{tabular}
\end{table}

We observed that for the raw URL group, out of the 163 users who opened the email using a computer email client, 8.0\% of them clicked, and for those who opened it with mobile devices, out of 233 users, 7.7\% clicked; these two click rates are reasonably close. However, we observed a difference when the phishing link is hidden from the user. For the button group, 10.9\% out of the 138 computer users clicked on the link, whereas for mobile device users, 20.2\% out of 233 users clicked. A similar difference can be observed for the hypertext group, out of the 325 computer users, 18.5\% of them clicked on the link, and out of the 260 mobile device users, 26.9\% of them clicked. These results add to our findings above, that link presentation interacts with device type to further shape users' behavior.

\medskip

% Lastly, we used two chi-squared tests to examine how users behavior in the three conditions differ when using computers and smartphones/tablets. For computers, the relationship between the three conditions are significant (p = .003), and the effect size is .13, this means for computer users, the phishing link presentation has significant and small effect on users clicking behavior. Compared to phishing emails with raw phishing URL, users are 1.4 time more likely to click on links that display as buttons, and 2.3 time more likely when masked with text. Similarly, for smartphones/tablets, the relationship is also significant (p $<$ .001), with an effect size of 0.20. This means phishing link presentation has significant and small effect on users clicking behavior, and this effect is greater than computer users. Users are 2.6 time more likely to click on the button version, and 3.5 time more likely on the hypertext version.

% TODO: do a comparison for smartphone link, etc.

\subsection{Link Presentation}

We further categorized our data into three groups based on the specific phishing email users received. During the campaign, we also observed that for the button version, the Windows Outlook application would display the button as hypertext. From the user's perspective, they are seeing the hypertext version, which is the same as if the user is been grouped in the hypertext group. Therefore, we decided to move the Windows Outlook application data from the button version, and add it to the hypertext version. Table \ref{tab:campaign1_separated} shows the number of users who opened and clicked on the phishing link using computers and mobile devices in each condition. Based on these data, we performed a series of chi-squared tests, including tests within the group between devices, and users clicking behavior between groups, as shown in Table \ref{tab:campaign1_all}.

\begin{table}[h]
\centering
\caption{Distribution of the campaign users' clicking actions using different devices}
\label{tab:campaign1_separated}
\begin{tabular}{|p{1.5cm}|>{\raggedleft\arraybackslash}p{1.5cm}|>{\raggedleft\arraybackslash}p{1.3cm}|>{\raggedleft\arraybackslash}p{1.6cm}|r|}
\hline
                          & \textbf{Computer}  & \textbf{Mobile}  & \textbf{Unclassified}   & \textbf{Sum} \\ \hline
\textbf{Raw URL clicked}         & 13                 & 18              & 11          & 42           \\ \hline
\textbf{Raw URL not clicked}     & 150                & 215             & 377          & 742          \\ \hline\hline
\textbf{Button clicked}          & 15                & 47               & 29         & 91           \\ \hline
\textbf{Button not clicked}      & 123               & 186              & 401         & 710          \\ \hline\hline
\textbf{Hypertext clicked}     & 60                & 70               & 22         & 152           \\ \hline
\textbf{Hypertext not clicked} & 265               & 190              & 422         & 877          \\ \hline\hline
\textbf{Sum}                     & 626               & 726              & 1262         & 2614         \\ \hline
\end{tabular}
\end{table}

\begin{table}[h]
\centering
\caption{Click behavior over all platforms: chi-squared test results; residual shown for first condition shown}
\label{tab:campaign1_all}
\begin{tabular}{|p{2.2cm}|>{\raggedleft\arraybackslash}p{1.2cm}|>{\raggedleft\arraybackslash}p{1cm}|>{\raggedleft\arraybackslash}p{1.3cm}|}
\hline
\textbf{Conditions}              & \textbf{$\chi^2(1)$ p-value}  & \textbf{Effect size $\phi$} & Pearson Residual for clicks\\ \hline
Masked vs. Raw URL       & $<$ .001      & .12         & +3.1\\ \hline
Button vs. Raw URL              & $<$ .001      & .10     &  +2.8 \\ \hline
Hypertext vs. Raw URL          & $<$ .001      & .15        & +4.0 \\\hline \hline
Hypertext vs Button          & .02           &   .05      &  +1.4 \\\hline
\end{tabular}
\end{table}

\subsubsection{H2 Testing}

%\textbf{Robt: WE SHOULD MAYBE SHOW THE CONTINGENCY TABLES FOR THE COMBINED MASK TESTS, THOSE TESTS ARE IMPORTANT< AND IF I WAS A REVIEWER, ID WANT TO SEE THEM.}

% raw vs. masked text
Hypothesis H2 stated: \textit{Users are more likely to click on phishing links when masked with buttons or hypertext than when shown as URLs.}

To test H2, we combined the button group and hypertext group into a combined masked text group, and compared it with the raw URL group, across all platforms, as shown in Table \ref{tab:raw_vs_maskedText}. We then performed a chi-squared test, and also calculated the effect size, with result: $\chi^2 (1,  N = 2614) = 34.8, p < .001, \phi = .12$ (Pearson residual: $+3.1$). H2 is thus supported, and with a small to moderate effect size. We therefore conclude that there is a difference in user behavior when seeing a raw link vs. a link concealed by a button or hypertext. 

\begin{table}[h]
\centering
\caption{Contingency table for raw URL vs. masked text}
\label{tab:raw_vs_maskedText}
\begin{tabular}{|p{1.8cm}|>{\raggedleft\arraybackslash}p{1.5cm}|>{\raggedleft\arraybackslash}p{1.5cm}|>{\raggedleft\arraybackslash}p{1cm}|}

\hline
                     & \textbf{Raw URL} & \textbf{Masked URL} & \textbf{Sum} \\ \hline
\textbf{clicked}     & 42               & 243                  & 285          \\ \hline
\textbf{not clicked} & 742              & 1587                 & 2329         \\ \hline
\textbf{Sum}         & 784              & 1830                 & 2614         \\ \hline
\end{tabular}
\end{table}

The interpretation of the effect size of .12 is small to moderate, but we note that the difference is between the raw link case of 42 clicked vs. 742 non-clicked (5\% of total) and the masked link case of 152 vs. 877 (14\% of total). In practice, this could be a substantial difference in phishing susceptibility. 

%In our data, none of the chi-squared tests have a large effect size. However, for a relation to be considered as having at least a medium effect size, the differences between groups has to be huge. We believe that even though the effect size of the relations is small in our study, it is still noticeable and worth studying. 

% raw vs. button, and raw vs. hypertext

We also performed a chi-squared test between each of the two masked text variations, the button version, and the hypertext version, with the raw URL version. Both resulted in significant results. For the button variation, $\chi^2 (1,  N = 1585) = 17.4, p < .001, \phi = .10$ (Pearson residual: $=+2.8$),  and for the hypertext version, $\chi^2 (1,  N = 1813) = 41.27, p < .001, \phi = .15$ (Pearson residual: $=+4.0$).

\subsubsection{H3 Testing}

Hypothesis H3 stated: \textit{Users are more likely to click on phishing links if they are hidden with HTML buttons, compared with being hidden with hypertext.} 

To test H3, we compared the hypertext group data with the button group, across all platforms.  
We then performed a chi-squared test, and also calculated the effect size, with result: $\chi^2 (1,  N = 1830) = 5.2, p = .02, \phi = .05$ (Pearson residual: $+1.3$). The result is significant, but the residuals show the opposite to what we hypothesized. Users are slightly more likely to click on the hypertext version compared with the button version. H3 is thus not supported.

The above results are also shown in Table \ref{tab:campaign1_all}.

%  \textbf{GR: CORRECT ME IF I AM WRONG HERE, BUT IF YOU MOVED THE USERS THAT CLICKED ON OUTLOOK BUTTON VERSION OF THE EMAIL INTO THE HIDDEN HYPERTEXT GROUP, ARE YOU NOT SKEWING THE RESULTS TOWARDS THE HYPERTEXT?}

\subsection{Summary}

As shown in Table \ref{tab:hypothesisTable}, out of the three hypotheses we propose, all were significant at an alpha of $.05$, but only H2 had a non-negligible effect size. Users are more likely to click on the phishing link if it is displayed as a button or masked as text, compared with the raw phishing URL. We observed an important difference in users' behavior when the phishing link is been visually obscured. If the link is obvious, users' behavior is consistent regardless of the device. Whereas if the link is masked with a button or text, users are more likely to click on the link when using mobile devices than using computers. Among these two visual techniques, we hypothesized that the button version would make users more likely to click than the hypertext version, because of its affordance and visual appearance. However, in the study we observed a significant but opposite and very small effect. 

\begin{table}[thbp]
\centering
\caption{Summary of hypothesis testing}
\label{tab:hypothesisTable}
\begin{tabular}{|p{0.4cm}|p{5.5cm}|p{1.4cm}|}
\hline
& \textbf{Hypothesis statement}          &                                                                                                                                                                                       \textbf{Support} \\ \hline
H1  & Users are more likely to click on phishing links when using email clients on mobile devices, than when using computers.                                                                                                                   & Supported, but very small effect size\\ \hline

H2  & Users are more likely to click on phishing links when they are hidden using buttons or hypertext, than when shown as URLs.    & Supported, with small to moderate effect size\\ \hline
H3  & Users are more likely to click on phishing links if they are hidden with HTML buttons, compared with being hidden with hypertext.  & Not supported \\ \hline
\end{tabular}
\end{table}

\section{Discussion} \label{Discussion}

% discuss computer vs. smartphone. link back to the related work, that due to the physical limitations etc, or email client features, interaction methods, mindsets, users are less careful when using smartphones than computers.
% discuss apple mail

As pointed out in Steves et al.'s study\cite{steves2020categorizing}, ``not all cues are created equally". Different phishing cues have different levels of detection difficulty and effect. For instance, displaying phishing links as raw URLs is a very obvious cue that could help users determine that the email is phishing. Our study shows that regardless of the device people use for checking emails, they are less likely to click when seeing the raw URL version. The ease of identifying such a cue is partially because users are able to see the raw phishing link without any interaction with the interface (e.g., hovering). The same phishing cue can be much more difficult to identify if it is masked with a button or hypertext. When no obvious cues are visible to the users, they may process the email normally as if it is legitimate. And only when they see some cue, such as a link with an unfamiliar URL, might they become suspicious, might the change their mindset to be careful about potential phishing attacks. The extra step of expecting users to hover over the link is a danger. Our study found that more users clicked on the link when it was displayed as a button or hypertext. The process whereby suspicion is raised can be explained by the concept of the human mind as a dual process system \cite{kahneman2011thinking}, which has been applied in the domain of phishing susceptibility  \cite{vishwanath2018suspicion, wang2017coping, harrison2015examining, wright2014research}. This theory proposes that our mind has two reasoning systems, one is fast, effortless and based on heuristics, and the other is slow and requires systematic reasoning. When users see the raw phishing URL, their heuristic system is more likely to quickly raise a red flag and intuitively sense that something is wrong. This then triggers the systematic reasoning system to examine the email with more awareness, so they are less likely to click on the URL. Without visibility, Kahneman explains the conclusion is often WYSIATI: "What you see is all there is".
%In addition to Steves et al.'s study, we argue that not only do different phishing cues have different levels of detection difficulty and effect, different presentations of the same cue can also have different effects.

The use of different devices makes the problem of phishing presentation more complex. Our study reveals that mobile device users were more likely to click on the link than computer users. This is in line with the literature \cite{breitinger2020survey, mcgill2017old, bitton2018taxonomy, koyuncu2019security}. %, which suggests that mobile device users generally exhibit lower information security awareness. 
Moreover, our findings add nuance to the literature: phishing link presentation matters --- mobile users were not more susceptible for the raw URL but were for the masked URL. This suggests that the dangers of masked presentation are exacerbated for mobile users. Factors such as screen size, the interaction mechanics of viewing the full URL, and the external environment may contribute.

%We speculate that even when the users feel suspicious about the email, they are less likely to spend time and effort validating the legitimacy of the email. Instead, they may proceed to click on the link. Other factors such as screen size, interaction method and external environment may also contribute to the email reading process, which accounts for the data observed in our study. However, the nature of the study means that we are unable to narrow down the factors that influence users' decision-making. Based on our results, we argue that for phishing emails with obvious cues, users can easily detect the email regardless of the device used for reading, reducing the likelihood of clicking on the link. In contrast, when no obvious cues are visible to the user, the device used for reading email will contribute to their decision-making, which influences their susceptibility to phishing. Therefore, we conclude that the effect of phishing links' visual presentation is stronger than that of the devices being used.
% Besides, mobile device users are more vulnerable in dealing with phishing emails, thus requires more attention and targeted training to minimise the risk.

Detection of phishing cues may vary among individual users, as their prior knowledge and experience influence their identification of cues. Across organizations, users will have varying knowledge about URLs and web domains. This type of knowledge will shape users' ability to determine the legitimacy of links. 

It was anticipated that the button presentation would result in a higher number of clicks due to its affordance and visual design. However, we found that the hypertext presentation of the link elicited more clicks. The legitimate security notification sent by the organization also uses hypertext links for releasing quarantined emails, which may make hyperlinks more familiar and trustworthy. It is possible that buttons are typically utilized for actions that affect a current web page, while hyperlinks are used for redirecting users to a different web page. Since users are not expecting to release quarantined emails within the email itself, they may be less likely to expect a button compared to a hyperlink that redirects them to a website. It is also possible that the click rates of phishing buttons and hyperlinks may be context-specific and depend on the topic of the phishing message. 

The nature of the masking text may impact susceptibility. In our main study, we used 'Release held emails'. In our pilot study, we tested a different approach: we masked the phishing URL as a legitimate-looking URL from an internal domain that was familiar to all staff in the organization. We compared this approach with the raw URL version. This was piloted with 374 members of the IT service unit who were randomly divided into two groups for each variation. We observed a large difference in the number of user clicks between the two groups. The raw URL version resulted in a 3.8\% click rate (3 clicks out of 79 users who opened the email), whereas the masked URL version resulted in a 19.6\% click rate (11 clicks out of 56 users who opened the email). These users were from the IT service unit and were expected to have more experience in dealing with phishing emails, yet we still observed a substantial difference in the click rate. Our pilot study suggests that masking with a legitimate URL would be dangerous and would persuade users to click on the link. We chose not to include this 'legitimate URL' variation in the main campaign as we are focusing on generic phishing. Knowing the internal legitimate domains that the users are trusted and familiar with would be considered as tailored phishing, which is outside the scope of the current paper. In addition, both we and the IT service unit thought this variation was too difficult for the general population. We will consider this in future campaigns.
% \textbf{GR: MAYBE WE SHOULD SHOW THE EMAIL OR THE URL OR GIVE MORE DETAILS WHY IS LEGITIMATE (SAY THAT THE URL IS USED IN THE ORGANISATION AND SHOULD BE KNOWN TO THE IT DEPARTMENT. THE EXPLANATION ABOUT WHY WE DID NOT USE FOR THE GENERAL POPULACE IS RATHER WEAK.MAYBE LINK TO THE NIST LEVEL THAT WE AGREED TO TEST WITH THE IT SEC DEPART.}

% discuss the potential in crafting email targeting specific email clients
The presentation of the same phishing email can vary depending on the email client used. We observed that the button (created using a button tag inside a hyperlink tag) is displayed as a normal hypertext in Windows Outlook client. Windows Outlook client can display buttons normally through other methods, such as modifying the CSS style of the hyperlink. If hyperlinks are more effective in attracting clicks compared to buttons, as we found in our study, this means that Windows Outlook users may be more susceptible than users of other clients, as they may see more hyperlinks rather than buttons. 

%Attackers may be able to exploit differences between email clients to create targeted phishing emails. For instance, Windows Outlook client uses conditional CSS (with mso tag) to display content that only shows in Windows Outlook client. This means attackers can craft content that is more attractive for Windows Outlook client users to motivate clicks. 

Masked links increase susceptibility, and the display of those masked links differs by device and by platform. When computer web mail users hover over the link, the landing page URL appears in the bottom left corner of the browser. In contrast, when computer application users hover, the landing page URL generally would be displayed next to the mouse cursor. In some email clients, such as Outlook, the URL also appears on the bottom left corner of the application. The placement of URLs may impact the readability of the URL and cause a difference in the click rate. Moreover, hyperlink tags in HTML can be configured to include a tooltip with customizable content. While the tooltip attribute generally has no effect on the presentation of links in email client applications, it can mimic a hover box for web clients when a link is hovered. Attackers can exploit this technique to display the ``real landing page URL" next to the phishing link, thereby tricking web client users into believing the link is genuine. In an extreme case, a web client may display three different URLs when hovering over a single link: the link is visually displayed as one link, then the tooltip displays a second link, and the browser displays the real URL (could be a third URL) at the bottom left corner. 

Users' susceptibility to phishing can be reduced when the raw URL is easily visible. However, presenting raw URLs sacrifices usability for security. HTML rendering can make the text easier to understand, and can make the purpose of links clear, but it can also make text and links more deceptive. In particular, allowing links to be concealed is a gift to attackers.
%We strongly encourage future research to study the relationship between visual presentation of emails and email clients, as well as the potential for crafting phishing emails that target specific email clients. By examining the features of emails and email clients, we may gain insights into how to enhance the overall email communication security.

%One important question that has not been answered in our study is how many users have hovered over the phishing link before opening it. However, it is not possible to collect such data because the users are interacting with their own email inboxes, and 
Our research was not able to record hovering activities. It would be interesting to assess whether the users hovered over the link or not and how this impacts susceptibility. 
%We stress the necessity of find out how users read their emails, including how likely they are in hovering over links to check for legitimacy, and focusing on other important cues that help them identify phishing emails. 
Others have used eye trackers to examine where people look when reading emails \cite{pfeffel2019user}. Unfortunately, eye-trackers are not yet feasible for the scale of organization-level field studies. 
%It is crucial to understand how many users actually will hovering over links when reading, and how often they do so. Phishing training has been greatly focus on educating users to look for links \cite{jampen2020don}, but if we do not have the statistics, it may difficult to make improvement in the training to better help the users.

%We should explore ways to accomplish both where possible. 

The design elements in email clients deserve more attention in future phishing research. For example, email service providers such as Gmail \footnote{https://support.google.com/mail/answer/1311182?hl=en} hide the sender's email address when a user has replied to them. Given that most phishing emails are sent by unknown senders, highlighting the difference between known and unknown senders in the email client could serve as a helpful cue for identifying phishing emails. In light of the results of our study, a potential solution could be to display the raw URLs when the sender is unknown. Future research could manipulate design elements in email clients to demonstrate the potential in helping users protect themselves from phishing.

\subsection{Limitations}

The findings of our study may seem obvious, but our study provides evidence using a large-scale phishing campaign to validate that users are more susceptible to phishing when using mobile devices, and when seeing phishing links as masked text. This represents a danger that suggests a need for specific training or changes to email clients.

Our study has a number of limitations. First, the GoPhish framework uses invisible tracking images to determine email opening. If the email client does not allow the automatic downloading of pictures, this tracking method will fail. Therefore, we expect that we would have had more open attempts than are shown in our data. In our analysis, we discarded all the link clicks if they did not map to any open events. In reality, these click events were potentially legitimate user actions, but since we can not identify the corresponding open attempt, we can not determine the user-agent associated with the action. Hence this portion of the data is lost in our study.

Second, the classification of email clients and devices relies on the user-agent string, which is not always reliable. For instance, for Apple Mail users, by default, the user-agent associated with their actions would be ``Mozilla/5.0" across all devices, and the users can manually select which user-agent to be used when sending messages. We are pleased to see the benefit of using such an approach to protect users' privacy, but at the same time, it creates extra difficulties for us in verifying the applications and devices.

Third, in our study, we assume that none of the emails are forwarded. We acknowledge that there will be a small group of users who may automatically forward their emails to a different account, but we are unable to separate these users from the rest. For these users, the sender of the forwarded email will be the users themselves, which could potentially influence their behavior because users may pay less attention to the original sender of the email. We assume that these users would be evenly distributed in the three groups, so our results would not be affected.

Fourth, we should note that the URL used in our study was not specially crafted to resemble one that the users would recognize and trust. Instead, we chose one that attackers adopted or hijacked for general use. Domain names for phishing are now used and abandoned frequently, typically less than 24 hours, to avoid blocklisting \cite{bell2020}. We therefore suggest our URL is realistic for many attacks, though not representative of specialized and targeted campaigns.

Finally, in our study, we moved the Windows Outlook application users from the button version to the hypertext version because the Outlook Windows application consistently displays the phishing URL as hypertext in both groups. To ensure that this migration of data will not affect the findings we present, we also performed the same test with Windows Outlook application users being removed from all conditions and recomputed our tests to compare the results. As shown in Table \ref{tab:case_comp}, the results are very similar. 

Further investigations into this area will aid in the development of more robust solutions that can aid users in identifying and avoiding phishing scams. We encourage researchers to replicate our study and build upon the knowledge gathered to contribute to the ongoing efforts of improving online security.

\begin{table}[h]
\centering
%Windows Outlook impact analysis: Finding that in Windows Outlook (but not Mac Outlook) displayed the button condition email indistinguishable from hypertext, we explored two cases for resolution. 
\caption{
Comparison of hypothesis test results for the \textbf{Mov} case (moving Windows Outlook data from button to hypertext) and the \textbf{Del} case (deleting the Windows Outlook data from all conditions).}
\label{tab:case_comp}
\end{table}
% latex table generated in R 4.2.2 by xtable 1.8-4 package
% Fri Feb 17 17:51:20 2023
\begin{table}[ht]
\centering
\begin{tabular}{|l|p{2cm}|r|r|c|r|}
  \hline
\bf Case & \bf Hypothesis & $\chi^2$ & \bf p value & $\phi$ & \bf resi \\ 
  \hline
Mov & H1: all & 5.0 & .025 & .06 & 1.39 \\ 
  Del & H1: all & 6.5 & .011 & .08 & 1.32 \\ 
   \hline
Mov & H1: raw only & $<$.01 & 0.927 & $<$.01 & -0.06 \\ 
  Del & H1: raw only & .6 & .424 & .04 & 0.39 \\ 
   \hline
Mov & H1: button only & 5.4 & .02 & .12 & 1.29 \\ 
  Del & H1: button only & 5.4 & .02 & .12 & 1.29 \\ 
   \hline
Mov & H1: hypertext only & 6.0 & .014 & .10 & 1.61 \\ 
  Del & H1: hypertext only & 3.1 & .08 & .09 & .86 \\ 
   \hline
Mov & H2: masked vs. raw & 34.8 & $<$.001 & .12 & 3.05 \\ 
  Del & H2: masked vs. raw & 32.8 & $<$.001 & .12 & 2.98 \\ 
   \hline
Mov & H2: button vs. raw & 17.4 & $<$.001 & .10 & 2.81 \\ 
  Del & H2: button vs. raw & 20.6 & $<$.001 & .12 & 2.97 \\ 
   \hline
Mov & H2: hypertext vs. raw & 41.3 & $<$.001 & .15 & 3.99 \\ 
  Del & H2: hypertext vs. raw & 36.6 & $<$.001 & .15 & 3.89 \\ 
   \hline
Mov & H3: button vs. hypertext & 5.2 & .023 & .05 & 1.40 \\ 
  Del & H3: button vs. hypertext & 2.8 & .092 & .04 & 1.10 \\ 
   \hline
\end{tabular}
\end{table}

\section{Conclusions}  \label{Conclusion}

In this paper we presented the results of a phishing campaign study conducted in a large organization. Our study focuses on the visual presentation of emails, and how it influences users' behavior. Our result shows that users are somewhat more susceptible to phishing when using mobile devices, compared to using computers. More importantly, we found that masking phishing links with buttons or hypertext has a greater impact in persuading users to click on the phishing link, as opposed to displaying the raw URL. With these masked URL conditions, mobile users were more susceptible compared to computer users. Our study identified novel link presentation variables that contribute to users' susceptibility to phishing. We recommend that future research further explore the presentation of phishing emails and how email clients can be designed to assist them in protecting themselves.

%Our study identified new variables that contribute to users' susceptibility to phishing. %Although the end users are the last line of defence against phishing attacks, they may not always be prepared for phishing emails. Hence, finding solutions that could help users identify phishing cues easily is key to reduce users' susceptibility to phishing. Future research should place greater emphasis on what users see when reading emails, and how email clients can be designed to assist them in protecting themselves.

%\textbf{GR: MAYBE WE SHOULD ALSO SAY THAT THE CUES SHOULD BE MORE AGNOSTIC OF THE DIFFERENT CLIENTS/PLATFORMS.} RObt: I added this point easier in the implications section.
%-------------------------------------------------------------------------------
\section*{Acknowledgments}
%-------------------------------------------------------------------------------

T.B.A.

%-------------------------------------------------------------------------------
\bibliographystyle{plain}
\bibliography{references}

%%%%%%%%%%%%%%%%%%%%%%%%%%%%%%%%%%%%%%%%%%%%%%%%%%%%%%%%%%%%%%%%%%%%%%%%%%%%%%%%
\end{document}